# Detect Related Bugs from Source Code Using Bug Information

Deqing Wang, Mengxiang Lin, Hui Zhang, Hongping Hu

State Key Laboratory of Software Development Environment, Beihang University,
No.37 Xueyuan Road, Haidian District, Beijing, 100191, P.R.China
{wangdeq, mxlin, hzhang, huhp}@nlsde.buaa.edu.cn

*Abstract*—**Open source projects often maintain open bug repositories during development and maintenance, and the reporters often point out straightly or implicitly the reasons why bugs occur when they submit them. The comments about a bug are very valuable for developers to locate and fix the bug. Meanwhile, it is very common in large software for programmers to override or overload some methods according to the same logic. If one method causes a bug, it is obvious that other overridden or overloaded methods *maybe* cause related or similar bugs.**

**In this paper, we propose and implement a tool Rebug-Detector, which detects related bugs using bug information and code features. Firstly, it extracts bug features from bug information in bug repositories; secondly, it locates bug methods from source code, and then extracts code features of bug methods; thirdly, it calculates similarities between each overridden or overloaded method and bug methods; lastly, it determines which method maybe causes potential related or similar bugs. We evaluate Rebug-Detector on an open source project: Apache Lucene-Java. Our tool totally detects 61 related bugs, including 21 real bugs and 10 suspected bugs, and it costs us about 15.5 minutes. The results show that bug features and code features extracted by our tool are useful to find real bugs in existing projects.**

## I.    INTRODUCTION

### A.    Motivation

Polymorphism is one essential feature of an object-oriented programming language, and overriding and overloading of methods are implementation of it in Java [1]. The definition of "override" in Javadoc [2] is: Indicates that a method declaration is intended to override a method declaration in a superclass. Therefore, many overridden or overloaded methods happen in large software projects written in Java. A simple example in Apache Lucene[1] is the method "equals", which totally appears 44 times in 43 distinct classes in source code of Lucene version 2.4.1. The code of "equals" in class RangeQuery is shown in Fig. 1. As we know, if two methods have the same method name, the two methods maybe have similar functions. Just as shown in Fig. 1 and Fig. 2, the two methods have the same method name, i.e. "equals", and the function of them is to compare this object for equality with another object. If one method causes a bug, it is obvious that other overridden or overloaded methods maybe cause related or similar bugs, because many overridden or overloaded methods are implemented according to the same logic. Just as the method "equals", we know the bug occurring in RangeQuery occurs in RangeFilter again.

Meanwhile, developers of open source software projects often maintain open bug repositories. In a bug repository, users can submit bugs that they encounter while using software, and developers can confirm and fix the submitted bugs in the next release. One bug in the bug repository usually contains bug identifier, bug description, bug comments, bug resolution, bug fixed version and so on. For example, just as shown in Fig. 3, one bug in Lucene-Java bug repository[2] is LUCENE-1587[3].

Is the information of a bug useful to us? The answer is definitely yes! From LUCENE-1587, what we can acquire is as following: LUCENE-1587 occurs in class RangeQuery; the method involved by LUCENE-1587 is equals (), which detailed implementation is shown in Fig. 1; the reason causing LUCENE-1587 is that equals () does not compare collator property fully, i.e. when this.collator == null and other.collator != null, equals() should return false rather than true, so the bug occurs. The bug has been fixed in Lucene version 2.9.

Another question is whether the information acquired from LUCENE-1587 can be used to detect related or similar bugs. The answer is absolutely yes. We implement a tool Rebug-Detector (Related Bug Detector) and it detects a related bug, LUCENE-2131, as shown in Fig. 2. We submitted it to Lucene bug tracking system, and the developers confirmed it. The reason for LUCENE-2131 is the same to LUCENE-1587, that is, the conditional statement in Line 313 in Fig. 2 is incomplete. Besides LUCENE-2131, our tool finds another related bug in class ConstantScoreRangeQuery. This related bug is not submitted, because we confirm that it has been fixed by checking source code of Lucene version 2.9.

---

```
Lucene2.4.1/org/apache/search/RangeQuery.java

226  public boolean equals(Object o) {

227    if (this == o) return true;

228    if (!(o instanceof RangeQuery)) return false;

229    final RangeQuery other = (RangeQuery) o;

230    if (this.getBoost() != other.getBoost()) return false;

231    if (this.inclusive != other.inclusive) return false;

232    if (this.collator != null && ! this.collator.equals(other.collator))  // bug

233        return false;

234    if (this.lowerTerm != null ? !this.lowerTerm.equals(other.lowerTerm) : other.lowerTerm != null)

235        return false;

236    if (this.upperTerm != null ? !this.upperTerm.equals(other.upperTerm) : other.upperTerm != null)

237        return false;

238    return true;

239  }
```

Figure 1.  The code of "equals" in RangeQuery of Lucene version 2.4.1, the method compares this RangeQuery for equality with another object. The conditional statement (line 232) is false when this.collator == null and other.collator != null, so line 233 is not executed. In such case, the method should returns false rather than true, so LUCENE-1587 happened.

```
Lucene2.4.1/org/apache/search/RangeFilter.java

306  public boolean equals(Object o) {

307    if (this == o) return true;

308    if (!(o instanceof RangeFilter)) return false;

309    RangeFilter other = (RangeFilter) o;

310    if (!this.fieldName.equals(other.fieldName)

311        || this.includeLower != other.includeLower

312        || this.includeUpper != other.includeUpper

313        || (this.collator != null && ! this.collator.equals(other.collator))   // bug

314        ) { return false; }

315    if (this.lowerTerm != null ? !this.lowerTerm.equals(other.lowerTerm) : other.lowerTerm != null)

316        return false;

317    if (this.upperTerm != null ? !this.upperTerm.equals(other.upperTerm) : other.upperTerm != null)

318        return false;

319    return true;

320  }
```

Figure 2.  The code of "equals" in RangeFilter of Lucene version 2.4.1, the method compares this RangeFilter for equality with another object. The conditional statement (line 313) is incomplete, and it causes a related or similar bug, just like LUCENE-1587.

```xml
<item>
    <title>[LUCENE-1587] RangeQuery equals method does not compare collator property fully</title>
    <link>http://issues.apache.org/jira/browse/LUCENE-1587</link>
    <description>The equals method in the range query has the collator comparison implemented as:<br/> (this.collator != null && !
    this.collator.equals(other.collator)) <p>When <em>this.collator = null</em> and <em>other.collator = someCollator</em> this method will
    incorrectly assume they are equal. </p> <p>So adding something like</p> <table class='confluenceTable'><tbody> <tr> <th class='confluenceTh'>
    (this.collator == null && other.collator != null)<br/> would fix the problem</th> </tr> </tbody></table></description>
    <environment />
    <key id="12422124">LUCENE-1587</key>
    <summary>RangeQuery equals method does not compare collator property fully</summary>
    <type id="1" iconUrl="http://issues.apache.org/jira/images/icons/bug.gif">Bug</type>
    <priority id="4" iconUrl="http://issues.apache.org/jira/images/icons/priority_minor.gif">Minor</priority>
    <status id="6" iconUrl="http://issues.apache.org/jira/images/icons/status_closed.gif">Closed</status>
    <resolution id="1">Fixed</resolution>
    <assignee username="markrmiller@gmail.com">Mark Miller</assignee>
    <reporter username="mplatvoet">Mark Platvoet</reporter>
    <created>Mon, 6 Apr 2009 14:33:21 +0000 (UTC)</created>
    <updated>Fri, 25 Sep 2009 16:23:21 +0000 (UTC)</updated>
    <version>2.4.1</version>
    <fixVersion>2.9</fixVersion>
    <component>Search</component>
    <due />
    <votes>0</votes>
-   <comments>
        <comment id="12697694" author="markrmiller@gmail.com" created="Fri, 10 Apr 2009 00:42:06 +0000 (UTC)">Thanks Mark!</comment>
    </comments>
+   <attachments>
    <subtasks />
+   <customfields>
</item>
```

Figure 3. The partial xml file of LUCENE-1587.

## B. Challenges

So how to acquire these features from bug information and use them to detect related bugs from source code are the main challenges. There are three detailed challenges. The first challenge is how to analyze bug information using Natural Language Processing (NLP) techniques. As we know, it is extremely difficult to automatically analyze bug information by NLP. Unlike news articles or API documentation, bug information is short and impossible to "understand" automatically, even with the most advanced NLP techniques [14]. To make things worse, bug information is usually not well written and not grammatically correct. Moreover, many programming language words mix with natural language words. For example, "collator" and "null", extracted from LUCENE-1587, have program domain specific meanings comparing to general dictionaries, and these words are important to help us find related bugs.

The second challenge is how to detect related bugs. From analyzing bug information, we can generally locate methods where bugs occur, but how to determine whether overridden or overloaded methods would cause related bugs is still difficult to deal with. The former static analysis tools focus on finding man-made mistakes in copy-pasted code or finding frequent sequences in large source code, such as CR-Miner [4] and PR-Miner [8], but they could not detect such bugs caused by logical rule violations.

Despite the above challenges, we must ensure our tool is feasible, efficient and effective in large software projects, so it should have the following properties: (1) accuracy: the detected related bugs need to be reasonably accurate, because too many false positives would waste developers more time; (2) usability: the tool is implemented for developers to detect related bug from large software projects, so it should be easy to use; (3) scalability: the tool should be scalable to handle large software projects with more than 200 class files and 100,000 lines of code.

## C. Our Contributions

In this paper, we propose and implement a novel related bug detection tool named Rebug-Detector. The tool uses NLP techniques to automatically extract bug features (including classes, methods and some useful attributes causing errors) from bug information in bug repositories; and then it automatically locates, extracts code features of bug methods from source code; lastly, it detects related or similar violations with little effort by comparing overridden or overloaded method features extracted from source code to bug features and code features. More specifically, this paper makes the following two contributions:

(1) We propose a general method to automatically extract bug features from bug information and code features from source code. Firstly, we obtain all xml files of bugs from bug repository websites, and then we extract the title, description, comments, version etc. Secondly, we use information extraction techniques to implement bug features extraction. Lastly, we use some mining technique to extract code features from source code, which maybe cause bugs.

(2) We propose an efficient and effective algorithm to detect related or similar bugs from large open source projects. We extract all the overridden or overloaded methods of each bug method, and then calculate the similarities between them and each bug method to determine whether they maybe cause related or similar bugs. For example, it just takes us about 15.5 minutes to detect 61 related bugs (including 21 real bugs and 10 suspected bugs) from Lucene version 2.4.1 with 330 class files and 106,754 LOC (lines of code). We confirm each related bug by checking source code of Lucene version 2.9 or reporting it to developers through bug tracking system. We find most of these bugs are semantic bugs that violate logical rules and are thereby difficult for previous tools [3, 4, 5, 6, 7, 8] to detect.

The rest of this paper is organized as follows. Section II briefly presents related work. Section III presents how Rebug-Detector automatically extracts bug features from bug information and code features from source code, and detects related or similar bugs. Section IV presents the evaluation results by using Rebug-Detector, and gives a case study and some discussions. And last Section V concludes the paper.

## II. RELATED WORK

We briefly describe the recent related work. Xie *et al.* [19] pointed out: "To improve both software productivity and quality, software engineers are increasingly applying data mining algorithms to various software engineering tasks." So far, many data mining and NLP techniques have been used to help programming developers and testers locate and detect bugs in large software. There are three types of usages in detecting bugs by using data mining and NLP techniques: usage in analyzing software requirement documents; usage in parsing code comments or API documentations; and usage in parsing source code.

The first type is to use NLP and data mining techniques to find violations in software documentations. For example, Kof *et al.* [9] used part-of-speech (POS) tagging to identify missing objects and actions in requirement documents. Sampaio *et al.* [13] created a technique for automatically identifying aspects in requirements, called EA-Miner. Baniassad *et al.* [10] investigated how to create links from design-level documents to the corresponding design patterns in code by using semi-automated NLP methods. Fantechi *et al.* [11] exploited syntactic parsing to analyze uses cases from requirement documents.

The second type is to use NLP and data mining techniques to analyze code comments or Application Programming Interface (API) documentations written in natural language corresponding software, and it can help us to find bugs in source code which are inconsistent with comments or API documentations. Tan *et al.* [5] leveraged NLP techniques to analyze code comments in source code and detected bugs which were inconsistent with comments or bad comments which were inconsistent with correct code. Zhong *et al.* [7] used NLP techniques to parse API documentations and inferred specifications which were useful to detect real bugs in existing projects.

The last type is to use NLP and data mining techniques to parse source code and find bugs. Fry *et al.* [3] implemented an automatically extractor which extracted verb–DO pairs from Java source code, and used verb-DO pairs to detect violations. Shepherd *et al.* [12] used various NLP techniques such as stemming and POS tagging to locate and understand action-oriented concerns. Li *et al.* [4] implemented a tool CP-Miner for detecting related bugs in operating system code, caused by copy-pasted code, which was popular in much large scalable software. Li *et al.* [8] also proposed a general method called PR-Miner, which used frequent itemset mining to efficiently extract implicit programming rules from large software code written in an industrial programming language, and automatically detected violations to the extracted programming rules. Lo *et al.* [6] proposed a technique to mine temporal rules from program execution traces, which could guide developers to understand program behaviors, and facilitate all downstream applications such as verification and debugging. Williams *et al.* [22] used project histories to improve existing bug finding tools. Kim *et al.* [23] implemented BugMem to find bugs by analyzing the history of bug fixes. Wang *et al.* [20] presented a new approach to examine whether a new bug report was a duplicate of an existing bug report using natural language and execution information.

Therefore, we think data mining and NLP techniques can help us understand source code and detect bugs in large software projects, and they can save the time and energy of developers.

These related works focus on source code, code comments or API documentations. As we known, reporters often point out straightly or implicitly error code where bugs occur when they submit them, so bug information is valuable for developers and testers to locate and fix bugs. In large software projects, many potential related bugs, which are same or similar to reported bugs in bug repositories, are hidden in other places. To the best of our knowledge, our work is the first step to leverage both bug information and code features to detect related bugs from source code. Our tool can help programming developers locate and detect some new related bugs, which can affect software stability if they are not fixed.

## III. REBUG-DETECTOR

Rebug-Detector has two major functionalities: automatically extracting bug features from bug information and code features from source code; detecting potential related bugs by calculating similarities between overridden or overloaded methods and bug methods according to features. We give an overview of Rebug-Detector in subsection *A*, describe how to extract bug features from bug information in bug repositories in subsection *B*, present how to extract code features from source code in subsection *C*, and describe how to calculate similarities between overridden or overloaded methods and bug methods in subsection *D*.

### A. Overview

Algorithm 1 presents the pseudocode of Rebug-Detector. It first parses xml files of bugs and extracts all class names and method names from source code (Line 1). Then it extracts bug features by analyzing bug information and extracts method code causing bug (the method is called ***bug method***) from source code (Line 3). And then it extracts all the overridden or overloaded methods of each bug method from source code (Line 5). It last calculates similarities between all overridden or overloaded methods and each bug method, and returns the method which similarity is bigger than threshold $\theta$ (Lines 6-8).

For example, we can extract the following information from LUCENE-1587 xml file (as shown in Fig. 3.): the method name is "equals", denoted by M@equals; the class name is "RangeQuery", denoted by C@RangeQuery; the component in project is "Search", denoted by Com@Search; the key attributes are "collator", "if (this.collator!=null&&! this.collator.equals(other.collator))" and "incorrectly", denoted by A@{collator = 3, if (this.collator != null && ! this.collator.equals(other.collator)) = 1, incorrectly = 1}, the number is the frequency of a key word or sentence, just like term frequency in Vector Space Model (VSM). We can acquire these key attributes by comparing them to code of equls() in class RangeQuery. Because "collator" is a key word in method equals, so we should improve the importance of "collator" when we extract key attributes from bug information. We call the above information as bug features, namely these features are extracted from bug information.

---

**Algorithm 1.** The pseudocode of Rebug-Detector

---

**Input:** xml files of bugs X = $\{x_1, x_2, ..., x_n\}$

    source code S

**Output:** all methods causing related bugs

1: Init: parsing X;

    extracting all class names and method names from S;

2: for each $x_i$ in X do

3:    BF = bug_features_extractor ( $x_i$ );

4:    for each method $m_i$ in BF do

5:      M( $m_{i1}, m_{i2}, ..., m_{it}$ ) = methods_extractor( $m_i$,S);

6:      Sims( $m_{i1}, m_{i2}, ..., m_{it}$ ) = similarity_calculation( $m_i$ );

7:      if (sim( $m_a, m_i$ ) > $\theta$)

8:        return $m_a$ ;

---

After obtaining above bug features, Rebug-Detector parses source code and extracts all the overridden or overloaded methods of M@methodname. For example, we find "equals()" appears 44 times in 43 distinct classes and "hashCode()" appears 42 times in 42 distinct classes by analyzing source code of Lucene 2.4.1. Because of the limitation of space, we do not show the detailed class names. If one bug has two or more bug methods, we express them as M@mehtodname1,… and M@methodnamen. For example, in LUCENE-1583 (SpanOrQuery skipTo() doesn't always move forwards), bug methods contain M@skipto, M@next and M@adjustTop etc. This case is difficult to deal with. In order to simplify such case, we will merge callees into the caller. In the above example, M@adjustTop is the callee, and M@skipto is the caller by analyzing the code of SpanOrQuery.skipto(). So we just parse M@skipto and ignore M@adjustTop.

The last step is to determine which method maybe causes a potential related bug. In object-oriented programming languages，polymorphism is a key feature. From one bug method, we can extract several dozen overridden or overloaded methods of the bug method from source code. Then we calculate the similarities between them and the bug method using similarity equation. If the similarity value is bigger than threshold $\theta$, which can be acquired heuristically, we think the method probably causes related bugs.

## B. Extracting Bug Features from Bug Information

The goal of bug features extractor is to automatically extract features from bug information in bug repositories, that is to say, to leverage bug information written in natural language by bug reporters and developers. They usually comment one bug and point out what causes the bug in natural language. The title, description and comments in bug files are valuable to us. To achieve this goal, we need to address two main challenges: (1) What to extract? (2) How to extract? This subsection presents our solutions to the two challenges.

Addressing the first challenge requires to consider the following issue. What type of information in bug repositories is useful to locate and detect bug? From analyzing bug information, we find two types of information can help developers to locate bug. The first type is the attributes written in natural language and the other type is the attributes written in programming language. For example, in LUCENE-1587, "RangeQuery equals method does not compare collator property fully." belongs to the first type, whereas"(this.collator!=null&&!this.collator.equals(other.collator))" and "(this.collator==null&& other.collator! = null)" belong to the second type. Generally, we can extract the class name, method name and some key attributes from the first type; and the second type can help us to locate programming statements. In order to improve the efficiency of bug features extractor and make our Rebug-Detector feasible in large open software projects, we define the following features about one bug, as shown in Table I.

Since we have known what features should be extracted, the next step is to solve how to extract such features. Here we leverage NLP techniques and feature extraction techniques. The process of bug information parsing and bug features extracting is shown in Fig. 4. It first uses JDOM to parse all xml files of bugs downloaded from Lucene-Java bug repository and then divides such information into sentences. We use "!", "?", ";" and "." together as sentence delimiters. The sentence separation task is importance because it involves correctly deciding which type the sentence belongs to. If the division is wrong, it will affect our features extraction. In addition, dot and exclamatory mark are frequently used in programming language, e.g., this.collator!=null. In such case, they should not be considered as sentence boundary. Furthermore, sometimes there is no any delimiter in the end of a sentence.

We define three types of features according to our requirements, they are: ***Special Features, Program Features*** and ***Natural Features***. The type of a sentence is determined when it is separated. To the first type, we just check whether one sentence contains special terminology by contrasting special terminology table (containing all class names and method names extracted from source code). To the second type, we must first locate and extract code of bug method, and then compare the sentence to it. The rest is the third type. If a sentence belongs to the first or the third type, word splitters [15] is used to split each sentence into words. Afterward, we use Part-of-Speech (POS) tagging [15, 16] to tell whether each word in a sentence is a verb, a noun, an adjective or an adverb. We just select the noun, the adjective and the adverb to denote a sentence. If a sentence belongs to the second type, we find common substrings by comparing it to code of bug method.

| Feature ID | Features | Denoted by |
|---|---|---|
| F1 | The class where one bug occurs | C@classname |
| F2 | The methods where one bug occurs | M@methodname1, M@methodname2 … |
| F3 | Attributes of programming language | P@text1, P@text2 … |
| F4 | Attributes of natural language | N@text1, N@text2 … |

| ID | RangeQuery.equals() | RangeFilter.equals() |
|---|---|---|
| 1 | if (this == o) return true  (freq = 1) | |
| 2 | if (!(o instanceof        (freq = 1) | |
| 3 | (this.Collator!=null&&!this.Collator.equals(other.Collator))    (freq = 1) | |
| 4 | if (this.String!=null ? !this. String .equals (other. String):other. String != null)  (freq = 2) | |
| 5 | return false  (freq = 3) | |
| 6 | return true   (freq = 1) | |

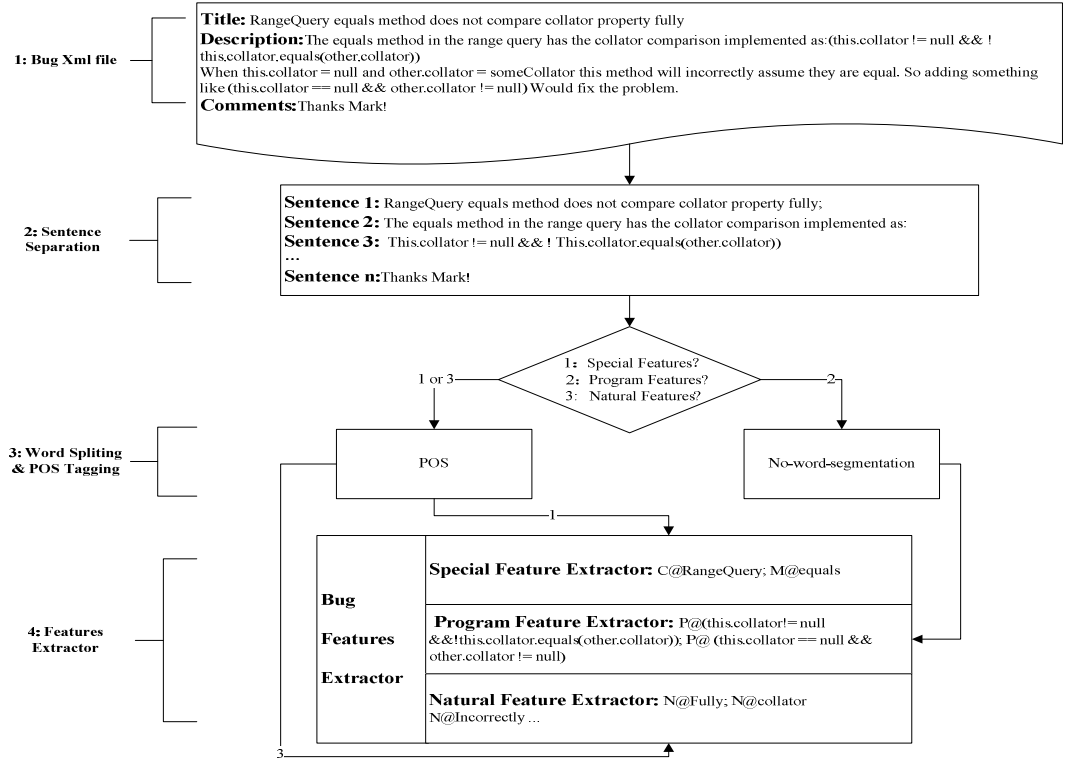

Figure 4.    The process of bug information parsing and features extracting: according to our requirement, Rebug-Detector divides features into three types, and extracts three types of features by three extractors, respectively.

## C. Extracting Code Features from Source Code

In this paper, we propose a novel solution to acquire code features of methods from source code. Our solution first parses and locates code of one method by scanning source code, and then replaces all the variables with their class names, and last extracts common substrings (CSs) from two pieces of code.

There are many techniques to finish the first phase. Two types of previous techniques are used to locate code. One type is based on string, in which the program is divided into strings (typically lines); the other type is based on token, in which the program is divided into a stream of tokens. In this paper, we consider the code of a class to be a text and use string-based to locate and extract the code of one method. We ignore all the blank lines and comments lines in the parse of source code.

Then, we replace all local variables with their class names according to the declarations of these variables. For example, the statement in method equals, "if (this.collator!= null && !this.collator.equals(other.collator))", is converted into "if(this.Collator!=null&& !this.Collator.equals(other.Collator r)) ", because the declaration of collator is "private Collator collator;", which can be extracted from source code using java reflection technique. This conversion will avoid that distinct variable names of the same type are seen as distinct strings.

The last phase is to find common substrings by Suffix Array (SA) [17, 18]. To find CSs of two strings, the time complexity of SA algorithm is linear. So the SA algorithm is effective and efficient for us to find CSs in two pieces of code. After we acquire all the CSs, we next consider these CSs to be code features, which will be used to calculate similarity of two methods. For example, the code features of RangeQuery.equals() and RangeFilter.equals() are as shown in Table II.

## D. Calculating Similarity and Detecting Related Bugs

The code features extracted from two pieces of code are important to determine the similarity between them. Besides, the program features extracted from bug information have the same importance. Given two pieces of code, C1 and C2, common substrings of C1 and C2 is denoted by CSs. Program features are denoted by P@. Natural features are denoted by N@. We use the following equation to calculate the similarity between C1 and C2.

$$sim(C1,C2)=\begin{cases}1, \text{if } \exists t,(a\in P@\wedge a\in CSs)\\ \dfrac{2*(Num(CSs)+Num(SN@)-Num(CSs\wedge SN@))}{Num(C1)+Num(C2)}, \text{otherwise}\end{cases}$$

Where $sim(C1,C2)$ denotes the similarity between C1 and C2, $Num(CSs)$ is the accumulate of frequency of each common substring in CSs, $SN@$ denotes the sentences which contain the same words in N@, $Num(SN@)$ is the count of sentence in $SN@$, $Num(CSs\wedge SN@)$ is the count of common sentences in CSs and SN@, $Num(C1)$ and $Num(C2)$ are lines of code of C1 and C2, respectively.

If a substring exists in P@ and CSs, $sim(C1,C2)$ is assigned to 1. Because we think detecting a potential bug is more valuable than detecting a false positive. Otherwise, $sim(C1,C2)$ is calculated by the second equation. Here we give a detailed explanation about the second equation. We think N@ extracted from bug information is important for us to locate some attributes in a method. For example, the reporters often give some keywords standing for variables in a method when they submit a bug. If a statement in C1 and other statement in C2 contain the same words (one or more) in N@, we suppose the statements are related to the bug and one of them is added to $SN@$. Maybe there are some common sentences in CSs and $SN@$, so they are subtracted in the second equation. The equation is similar to *Similarity* defined in CloneDR [21].

For example, we obtain bug method ("equals") causing LUCEN-1587, and we extract 43 pieces of code from 42 distinct classes, denoted by EC = {C1, C2… C43}. Then we calculate similarity between bug method and each element of EC, and similarities are denoted by $sims=\{sim(1,equals),sim(2,equals)...sim(43,equals)\}$. If $sim(i,equals)>\theta$, $1 \le i \le 43$ where the threshold $\theta$ can be acquired heuristically, we will think the method Ci probably causes a related bug. Our tool will return Ci and highlight the statements, co-occurrence sentences in P@ and CSs. Developers can check the code and verify whether the related bug is a real bug.

## IV. EVALUATIONS

### A. Experiment Setup

We have evaluated Rebug-Detector on Apache Lucene-Java. The number of files (only java files), lines of code (LOC), and brief description etc are shown in Table III. In order to evaluate our tool and verify whether related bugs are real bugs, we obtain 26 fixed bugs, which were found in version 2.4.1, from Lucene Change Log of Release 2.9.0. We first obtain 26 fixed bug IDs from Lucene-Java bug repository, and then we download each bug as an xml file, named by bug ID. Finally we acquire 25 xml files, except LUCENE-1658, which xml file is incomplete. Through parsing xml files, we find LUCENE-1611 could not extract any method names, so we discard this file. Our training data set just leaves 24 bugs. 24 bugs involve 71 distinct methods. In our experiments, we run Rebug-Detector on an Intel Core(TM) 2 Duo CPU 2.53GHz machine with 2GB memory and Windows XP system.

We set the threshold $\theta$ used in Rebug-Detector as 50%, which can be obtained from heuristic experiments. In our threshold heuristic experiments, we set $\theta = 10\%$ at the beginning, and then $\theta$ increases by 10% until $\theta=100\%$. It is easy for us to understand. $\theta$ is smaller, false positive bugs in related bugs are more, vice versa. Considering the trade-off between efficiency and effectiveness, we set $\theta = 50\%$. In the next experiments, $\theta = 50\%$, by default.

TABLE III. SOFTWARE EVALUATED IN OUR EXPERIMENTS, INCLUDING THE SOFTWARE VERSION, NUMBER OF CLASS FILES, LINES OF CODE (LOC), LANGUAGE, DESCRIPTION AND BUG INFORMATION. VALID BUGS MEAN THE NUMBER OF BUGS WHERE WE CAN EXTRACT BUG METHOD NAMES. INVOLVED METHODS MEAN THE NUMBER OF DISTINCT METHODS WHICH APPEAR IN 24 BUGS.

| Software | Version | Number of class files | LOC | Language | Description | Bugs fixed in version 2.9.0 | Valid bugs | Involved distinct methods |
|---|---|---|---|---|---|---|---|---|
| Lucene -Java | 2.4.1 | 330 | 106,754 | Java | full- text search engine library | 26 | 24 | 71 |

TABLE IV. THE STATISTICAL RESULTS OF RELATED BUGS DETECTED BY REBUG-DETECTED, #BUG_METHODS IS THE NUMBER OF DISTINCT METHODS EXTRACTED FROM BUG INFORMATION; #NUMBER_OF_METHODS IS SUM OF EACH OVERRIDDEN OR OVERLOADED METHOD EXTRACTED FROM SOURCE CODE; #RELATED_BUGS IS THE NUMBER OF RELATED BUGS DETECTED BY OUR TOOL, WHICH INCLUDES REAL BUGS, SUSPECTED BUGS AND FALSE POSITIVE BUGS; TIME (SEC) STANDS FOR THE TIME USED TO FIND THESE RELATED BUGS.

| Bug ID | #Bug_method | #Number_of_ methods | #Related_Bugs | | | Time (sec) |
|---|---|---|---|---|---|---|
| | | | real bugs | suspected bugs | false positive | |
| LUCENE-1327 | 3 | 76 | 1 | 1 | 3 | 102 |
| LUCENE-1415 | 2 | 84 | 0 | 1 | 2 | 123 |
| LUCENE-1590 | 2 | 5 | 1 | 0 | 2 | 20 |
| LUCENE-1587 | 1 | 43 | 2 | 1 | 1 | 78 |
| LUCENE-1600 | 3 | 36 | 8 | 2 | 5 | 65 |
| LUCENE-1647 | 1 | 22 | 2 | 1 | 3 | 45 |
| LUCENE-1599 | 2 | 64 | 2 | 0 | 3 | 112 |
| LUCENE-1583 | 6 | 97 | 1 | 2 | 5 | 300 |
| LUCENE-1801 | 4 | 17 | 2 | 2 | 4 | 40 |
| LUCENE-1847 | 1 | 20 | 2 | 0 | 2 | 48 |
| Total | | | 21 | 10 | 30 | 933 |

## B. Experiment Results

We present the evaluation results of Rebug-Detector in this subsection, including the statistical results of related bugs in Lucene-Java and a case study. Our goal is to find related bugs from source code using bug information, so the best measure is the number of related bugs detected by Rebug-Detector. Besides, the number of false positive bugs and time of finding related bugs are also important to a scalable tool. As we know, if one tool produces too many false positive bugs, it will waste the time and energy of developers. In order to make our tool feasible in large software projects, we must ensure our tool is efficient, that is to say, the scalability of the tool.

**Detecting Related Bugs:** Because of the limitation of space, we just list bug identifiers from which we can acquire related or similar bugs. The detailed statistical results are shown in Table IV. Our tool Rebug-Detector has totally detected 61 related bugs, including 21 real bugs, 10 suspected bugs and 30 false positive bugs (invalid bugs) from 10 bug files, and it takes us about 15.5 minutes. In order to confirm whether a bug is a real bug, we have two manners. The first is to report the related bugs to developers by bug tracking system, such as LUCENE-2131. We reported this bug to Lucene bug tracking system and the developers have confirmed it as duplicate, but it is a real bug and the bug has been fixed in the new release. The second is just to analyse and check it by contrasting source code of the new release. If the source code has been changed, we think the related bug is a real bug. To the suspected bugs, we will report them to Lucene bug tracking system, and communicate with the developers.

Here we give a case study about the related bugs detected by our tool. For example, the bug (LUCENE-1600: Reduce usage of String.intern(), performance is terrible.) shows that the usage of intern() will bring down the performance of Lucene and ultimately affect usage of customers. We find intern () is called for 36 times in Lucene 2.4.1. With our tool, we found a suspected bug of intern () in the class Field. We checked Lucene's latest version and confirmed that this suspected bug was a real bug. The left code snippet of Fig. 5 shows the found suspected bug, and the right code snippet of Fig. 5 shows how the suspected bug is fixed.

Through analyzing code of related bugs, we find that most of related bugs are caused by the same reason. When developers of software projects write programming code, they often override or overload some methods according to the same logic. It is common to software programmers. But if one method causes a bug, it is obvious that other overridden or overloaded methods would cause the related or

```
public Field(String name, TokenStream tokenStream, TermVector termVector) {
    if (name == null)
        throw new NullPointerException("name cannot be null");
    if (tokenStream == null)
        throw new NullPointerException("tokenStream cannot be null");

    this.name = name.intern();        // field names are interned
    this.fieldsData = tokenStream;

    this.isStored = false;
    this.isCompressed = false;

    this.isIndexed = true;
    this.isTokenized = true;

    this.isBinary = false;

    setStoreTermVector(termVector);
}
```

```
public Field(String name, TokenStream tokenStream, TermVector termVector) {
    if (name == null)
        throw new NullPointerException("name cannot be null");
    if (tokenStream == null)
        throw new NullPointerException("tokenStream cannot be null");

    this.name = StringHelper.intern(name);     // field names are interned
    this.fieldsData = null;
    this.tokenStream = tokenStream;

    this.isStored = false;
    this.isCompressed = false;

    this.isIndexed = true;
    this.isTokenized = true;

    this.isBinary = false;

    setStoreTermVector(termVector);
}
```

Figure 5.   A confirmed bug about method intern.

similar bug, and our work verifies this point. These related bugs are difficult to detect by other former tools, which focus on sequence mining and detect bugs caused by man-made mistakes, such as forgetting to modify names of variables in copy-pasted code.

Our tool does not work for the rest 14 bug files. By analyzing these files, we found that there are two reasons. (1): extracting too many bug methods, such as extracting 36 bug methods from LUCENE-1593, it is difficult for our tool to determine which method maybe causes the bug; (2): some bug messages are produced by the programming language, such as LUCENE-1623(Back-compat break with non-ascii field names), which will produce IOException error if a field name contains non-ascii characters in an index. This case is difficult for our tool to deal with, because our tool does not consider errors produced by programming language. In future work, we will complete this case.

Meanwhile, our tool finds 30 false positive bugs. The false positive rate is around 49.2%, and it is high. These invalid bugs will waste programmers more time and energy. In the future work, we will improve our tool to reduce false positive bugs.

In summary, using bug information and code features, we find various bugs that are related to bugs in bug repositories. The results demonstrate the usefulness of our proposal because developers do produce much overridden code.

**Efficiency:** Just as shown in the right column of Table IV, we can conclude the time used to find related bugs is feasible. It just takes us about 15.5 minutes to find 61 related bugs from 106,754 LOC, including 66,034 non-comment LOC, and it includes the time of parsing xml file, extracting bug features, parsing source code, extracting common substrings and calculating similarity.

### C.  Comparisons with PMD

There are many existing bug finding tools, such as PMD [24], BugMem [23], JLint [25] and FindBugs [26]. In order to evaluate the effectiveness of Rebug-Detector, we compare our tool with PMD. The tool is chosen because it does not require annotations and requires only Java source code as input. JLint and FindBugs need some predefined rules, which are project-specific or language-specific. BugMem needs data extracted from CVS. It is difficult and expensive to define and extract these rules. We will compare them with our tool in future work.

Here we use some bugs detected by our tool to verify whether PMD can detect them. To PMD-4.2.5, the value of parameter "duplicate chunks larger than" is set as 50 and other parameters are set as default. PMD totally detected 61 duplicate codes from source code of Lucene-Java version 2.4.1. We checked them carefully and found none of the duplicate codes referred to our related bugs. We think PMD focuses on duplicate code and our tool does not just focuses on source code. Much bug information is introduced in our approach. So, Rebug-Detector can find related bugs which cannot be detected by PMD, vice versa. And our tool is a complement of bug finding tools.

### D.  Threats to Validity

The threat to external validity includes the completeness of bug information in true practice. Although we apply our tool on a bug repository of a popular open source project (Lucene-Java), our tool is only evaluated on 24 bugs of Lucene-Java 2.4.1, the reason for choosing these bugs is that we can verify the detected related bugs by contrasting the new release. The threat could be reduced by more evaluations on more subjects in future work. The threat to internal validity includes human factors for determining bugs. To reduce the threat, we inspect bugs carefully and contact developers to confirm these bugs. The threats could be further reduced by involving more experienced developers in future evaluations.

## V.  CONCLUSIONS

This paper proposes a novel method to find related bugs from source code using bug information and code features. It first extracts bug features from bug information in bug repositories; and then it locates and extracts code features of bug method from source code, and calculates similarities comparing all overridden or overloaded methods to bug method. The last step is to determine which method maybe causes potential related bug. Our tool Rebug-Detector has detected 61 related bugs from Apache Lucene-Java, including 21 real bugs and 10 suspected bugs. The results show that our tool is useful for developers to locate and fix related bugs.

Our technique for extracting bug features and code features is more general. By replacing the frontend parser, Rebug-Detector can be easily modified to work with programs written in other OO programming languages, such as C++. In future work, we will use Rebug-Detector to do more experiments on large open source projects to verify its efficiency and effectiveness.

### ACKNOWLEDGMENT

We thank the anonymous reviewers for their insightful comments. The research is supported by the 863 High-Tech Program under Grant No. 2007AA010403.